# JUEGO SERIO PARA FÍSICA COMO ESTRATEGIA DE APRENDIZAJE ACTIVA Y LÚDICA

Pacheco-González, Alberto
TecNM campus Chihuahua
División de Posgrado e Investigación
Av. Tecnológico 2909
Tel. (614) 201-2000
alberto.pg@chihuahua.tecnm.mx

**RESUMEN.**

Se **presenta el diseño de un juego serio como estrategia de aprendizaje activo y lúdico para asistir la enseñanza de la Física, específicamente el tema de caída libre de objetos, con los siguientes propósitos educativos, primero para aplicar un conjunto de estrategias didácticas enfocadas en promover el aprendizaje activo supervisado y la programación en vivo y segundo, proveer un juego serio para ilustar las nociones de cinemática y simular la dinámica de cuerpos rígidos usando un motor de videojuegos. Gracias al uso del motor de juegos y una metodología integral fue posible auxiliar a que los estudiantes de Física implementaran dicho juego serio durante un semestre de un curso introductorio de Física.**

**Palabras Clave:** Simulación, matematización, juegos serios, educación de la física, motores de videojuegos, programación en vivo, desarrollo ágil.

**ABSTRACT.**

**The design of a serious game is presented that served as an instrument to motivate and aid to Physics education using active and ludic learning, specifically the topic of free fall of objects, with diverse educational purposes, first to apply a set of teaching strategies focused on promoting supervised active learning and live programming and second, providing a serious game to illustrate the notions of kinematics and simulate the dynamics of rigid bodies using a video game engine. Thanks to the use of the game engine and a holistic methodology, it was possible to help Physics students implement said serious game during a semester of an introductory Physics course.**

## 1. PROBLEMÁTICA

Un serio problema en la educación actual de las Ciencias y específicamente, de la Física, es el creciente desinterés, bajo rendimiento académico y alta deserción de los estudiantes, donde el método tradicional de enseñanza resulta inefectivo para estos estudiantes [1]. Para atender este problema [2, 3] recomiendan el uso de estrategias de aprendizaje activo y aprendizaje lúdico, i.e. juegos serios y gamificación.

El presente trabajo describe el diseño y desarrollo de un juego serio capaz de simular y estimular el aprendizaje de las leyes de la Física involucradas en la caída libre de objetos, partiendo de un estado de reposo con los siguientes propósitos educativos: 1) para los estudiantes universitarios consiste en matematizar los principios cinemáticos vistos de manera teórica en un curso de Física siguiendo un conjunto de estrategias enfocadas en promo–ver el aprendizaje activo supervisado destacando además, que la mayoría dichos estudiantes no tienen experiencia en programa–ción; 2) en un futuro próximo se pretende utilizar dicho juego serio con estudiantes de Física de nivel medio superior para ilustar las nociones de cinemática y dinámica de cuerpos rígidos.

## 2. MARCO CONCEPTUAL
### 2.1. Matematización.

A lo largo de la historia, la Ciencia ha progresado de forma inequívoca y gradual gracias a la verificación experimental de modelos teóricos, producto de la matematización, misma que permite sentar las bases para resolver problemas mediante el modelado matemático. Para desarrollar dicha competencia entre los estudiantes se recomienda que aprendan a transformar pro–blemas reales en modelos matemáticos [4].

La Física originada en el siglo XVII con Galileo Galilei y Simon Stevin, alcanzando su logro supremo en la obra de Isaac Newton, siendo un ejemplo contundente de la matematización como descripción matemática y la mecanización del mundo físico, cuantificando aquellos procesos que se esten analizando. Es decir, se refiere a definir modelos donde "entran números y salen números" [5]. Para reducir la brecha entre el dominio matemático y el modelado de fenómenos físicos, [6-8] sugieren la incorporación de herramientas digitales para soportar la matematización, así como también el pensamiento lógico, analítico, algebraico, sistémico, crítico y computacional, más recientemente englobados dentro de las habilidades STEM.

Las herramientas digitales juegan un papel crucial en la mejora de la matematización dentro de la enseñanza de la Física, al cerrar la brecha existente entre el dominio matemático y la realidad física [8]. Dichas herramientas pueden ayudar a conectar diferentes representaciones dentro del dominio mate–mático, apoyando a los estudiantes a manipular las variables dentro de los modelos de la Física. Sin embargo, a pesar de los beneficios potenciales de las herramientas digitales, existe una curva de aprendizaje para que los educadores implementen con dichas herramientas dentro de las aulas.

## 2.2. Entornos, herramientas y motores de juego.

Para crear videojuegos, a diferencia de sólo utilizar librerías de gráficación y componentes de juego, como OpenGL, WebGL, DirectX, Metal, Box2D, Cocos2D, PyQt, PyGame, un entorno integrado de desarrollo (*IDE*) ofrece una plataforma abstracta y un arsenal de herramientas, editores, motores, gestores, librerías y componentes reusables para crear videojuegos. La arquitectura de un entorno integrado para juegos ofrece una parte visible y otra oculta. La parte externa interactua con el desarrollador y ofrece diversas herramientas, como el editor de escenas, niveles, scripts, propiedades, eventos, recursos, animaciones, sonidos y efectos (Fig. 1). La parte interna oculta es el motor de juego (*game engine*) compuesto por un controlador de fuerzas, objetos y colisiones sujetas a las leyes de la Física (*physics engine*), un gestor de las eventos y controles de interfaz de usuario, un manejador de tarjetas gráficas (GPUs), manejador de audio, gestores de escenas, niveles, recursos, errores y un interprete de scripts.

A su vez, existen diversos géneros de videojuegos: acción, estrategia, aventura, deportes, casuales, de plataforma, Arcade, masivos y de roles (RPG, MMO, MOBA), simulación y educa–tivo. De acuerdo con [9], en México los géneros de acción, aventura y casuales son los más populares. Entre los motores de juegos más populares se encuentran Unity, Unreal Engine, Game Maker Studio, Scratch y Godot Engine, entre otros.

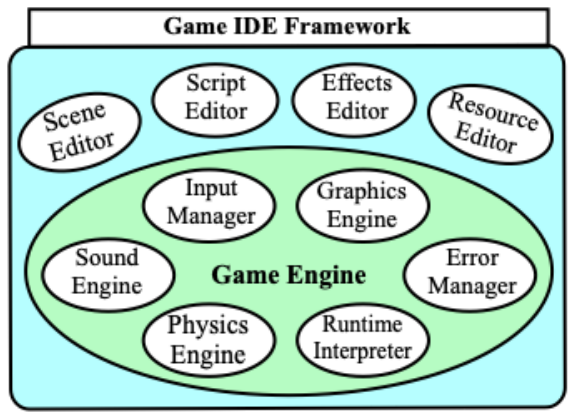

Figura 1. Principales subsistemas de un motor de juego.

El entorno Godot es un proyecto de software de código abierto, multiplataforma, bien soportado y documentado [10]. Este entorno ligero y bien integrado ofrece todos los compo–nentes básicos de un entorno (Fig. 1). Godot permite crear juegos 2D y 3D, soporta tarjetas gráficas basadas en Open GL y Vulkan, y posee un lenguaje de programación similar a Python llamado GDScript. Un juego se representa en Godot como un árbol de escenas y a su vez, cada escena es un árbol de nodos (Fig. 2). Existen además, muchos tipos de nodos: objetos cinemáticos, estáticos y rígidos, textos, botones, listas, ventanas, *timers*, *shaders*, imágenes, audios, animaciones, efectos, etc.

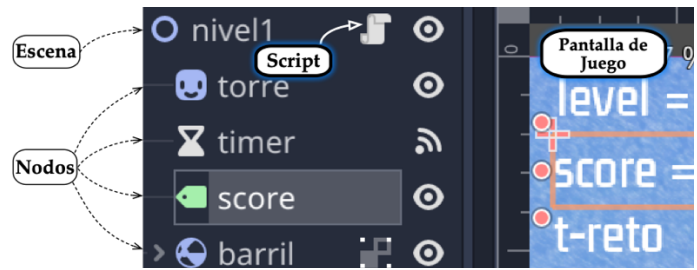

Figura 2. Entorno Godot: escena de un videojuego.

## 2.3. Juegos serios.

Un juegos serio es un videojuego que además de ofrecer es–parcimiento y diversión, integra intenciones pedagógicas, infor–mativas e interactivas orientadas a educar a un individuo [11]. Los juegos serios permiten fortalecer la creatividad, la resolución de problemas, la colaboración y participación de los estudiantes [2, 11-14]. Diversos estudios en torno al diseño de juegos serios enfatizan, sobre todo, la importancia de alinear la dinámica del juego con los objetivos de aprendizaje para soportar así, la adquisición del conocimiento por parte del estudiante [13, 14].

## 3. DESARROLLO

### 3.1. Metodología.

Para atender la problemática mencionada arriba, sobre el bajo interés y rendimiento de los estudiantes de Física, se desarrolló un juego serio como estrategia de aprendizaje activo y el apren–dizaje inductivo [15], misma que ha sido descrita a detalle en [16]. El juego atiende dos estrategias y objetivos educativos dis–tintos (Fig. 3): 1) para un jugador, el juego promueve un apren–dizaje inductivo y lúdico de la Física para descubrir ciertos los principios físicos de la caída libre; 2) para el estudiantes de Física o desarrollador del juego con la asistencia de un mentor (Fig. 3), su proyecto y las sesiones de programación en vivo sirvieron como evidencia de su aprendizaje y aplicación (matematiza–ción) de los temas de caída libre del curso de Física, conside-rando varias restricciones académicas, como: un tiempo limitado para realizar el proyecto, la falta de experiencia en programación y alto índice de reprobación de los estudiantes [16]. Por tanto, como resume la Fig. 3, la metodología fusiona tres flujos de trabajo concurrentes: $D_1$) para el estudiante-desarrollador, se aplican diversas estrategias didácticas, inclu–yendo un aprendizaje supervisado por un mentor [16]; $D_2$) para el jugador, como se mencionó, el juego promueve un aprendizaje inductivo y lúdico; $D_3$) para desarrollo rápido del juego serio se usó un motor de juego y un método ágil de prototipado en espiral, obteniendo de forma parcial y gradual varias versiones del juego [17, 18].

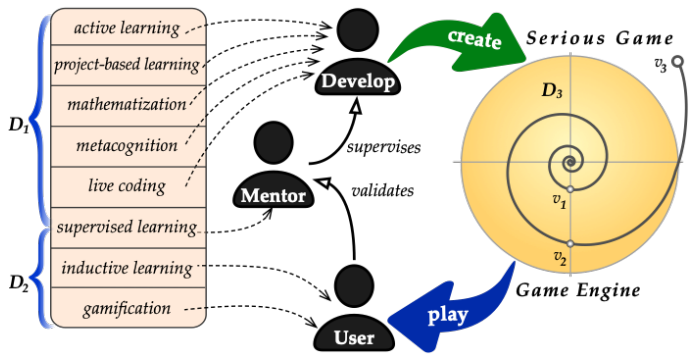

Figura 3. Metodología, agentes y flujos de trabajo.

### 3.2. Diseño y desarrollo del juego serio.

El juego consta de seis escenas principales (Fig. 4): la pantalla inicial, cuatro niveles de juego y la pantalla final con la puntuación total del juego. Se elaboró una planeación didáctica y un programa de trabajo para las sesiones semanales de programación en vivo para desarrollar los distintos módulos y versiones del juego como se detalla en la Tabla 1, con las siguientes columnas: 1) número de sesión de programación en vivo supervisada; 2) actividad y tema de Física considerado; 3) herramientas y componentes creados siguiendo la metodología de prototipado en espiral.

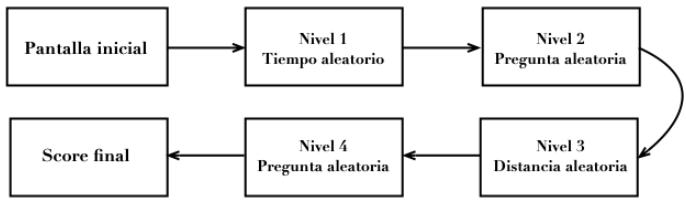

Figura 4. Pantallas y niveles del juego serio para Física.

El diseño de las pantallas de juego incluye dos plantillas superpuestas (*layouts*): 1) el tablero de juego (*heads-up display*, hud); 2) la escena del nivel actual. El tablero de juego (fig. 5) indica en todo momento el nivel de juego, reto por alcanzar, tiempo restante de juego y puntuación (*score*). De las sesiones 8 a 10 (Tabla 1) se desarrolló el tablero creando una escena tipo *singleton* (*autoload*) y varios nodos de control. Cada nivel de juego contiene (Fig. 6): una imagen de fondo, una regleta (escala de referencia), un personaje (objeto en caída libre), un peldaño móvil, una flecha, un barril y un piso. Cada estudiante desarrolló de forma libre y creativa sus propios recursos y características (aspecto, color, tamaño, *font*, ubicación, etc.) tanto del tablero como cada uno de los niveles de juego.

Tabla 1. Plan de trabajo: sesiones, temas y herramientas según [16].

| Sesión | Temas de Física incorporados | Herramientas de Godot aplicada en cada sesión |
|---|---|---|
| 1 | Investigar juegos para Física | Encuesta inicial diagnóstica |
| 2 | Historia de la Física (Galileo) | Simul. y matematización de la Física |
| 3 | Cuerpos estáticos y rígidos | Manejo del entorno, editores y ventanas |
| 4 | Caída libre y cte. gravedad | Manejo de recursos (assets, imágenes) |
| 5 | Propiedades: masa, posición, veloc. inicial, elasticidad | Manejo de propiedades (peso, giro, damping) |
| 6 | Torque angular, fricción, amortig. | Scripts, variables, funciones (rnd), eventos |
| 7 | Matematizar (desp, dim, prop, cte) | Detección de colisiones y control teclado (cursor) |
| 8 | Cálculo de tiempo | Tablero: hud y layouts |
| 9 | Actualizar tablero | Sonidos, timers y efectos |
| 10 | Cálculo de distancia | Manejo de niveles y var. globales |
| 11 | Mov. circular uniforme | UI: botón, lista, contenedor |
| 12 | Torque angular | Nivel 2: generador preguntas y opc. múltiples al azar |
| 13 | Deducción de segunda parte de modelo de caída libre | Nivel 3: Reusar nivel 1 con distancia al azar |
| 14 | Preguntas de Repaso para nivel 2 y nivel 4 | Pantalla de inicio. Nivel 4 (pregs). Pantalla final |
| 15 | Prueba del juego | Depuración de errores |
| 16 | Avances del juego de Física | Dudas y ajustes (código) |
| 17 | Reporte proyecto (parte 1) | Desc. pantallas y videos del juego |
| 18 | Jugadores invitados (nivel medio superior) | Encuestas finales |
| 19 | Reporte proyecto (parte 2) | Desarrollo y conclusiones |
| 20 | Presentación de juegos | Comentarios finales |

La dinámica de los principales elementos del juego ubicados en las pantallas de los niveles 1 y 3 son (Fig. 6): a) el personaje (objeto rígido) iniciando en reposo, al pulsar la tecla *enter*, el objeto cae libremente hasta impactarse en el piso o caer dentro de un barril; b) el usuario puede desplazar verticalmente un peldaño usando las teclas de cursor arriba/abajo; c) al inicio de cada jugada, un barril se desplaza horizontalmente a una posición fija de forma aleatoria; d) el usuario puede girar el peldaño usando las teclas izquierda/derecha para que, cuando impacte el objeto, el ángulo del peldaño establezca una trayectoria parabólica de tal forma que el objeto logre caer dentro del barril.

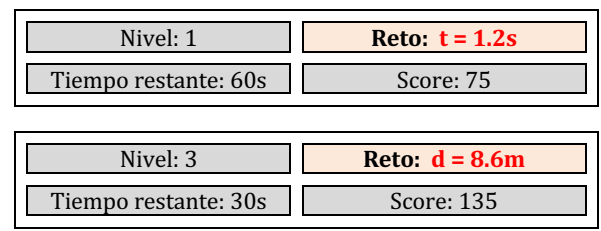

Figura 5. Tablero de juego para: a) nivel 1; b) nivel 3.

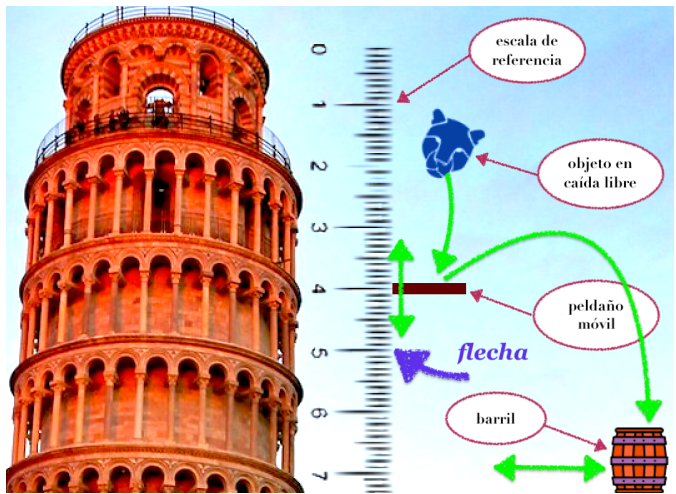

Figura 6. Principales elementos y controles del juego.

Las pantallas y escenas para el nivel 2 y nivel 4 son idénticas y ofrecen un cuestionario compuesto de controles UI Godot: contenedores, textos, iconos, listas y botones. A partir de un banco de preguntas correspondiente al nivel de juego anterior, se selecciona al azar y se presenta una pregunta y cuatro posibles respuestas (opciones) sobre los temas de cinemática, movi–miento uniformemente acelerado y caída libre de cuerpos físicos ilustrados y aplicados en la dinámica del juego. El jugador elige una de las opciones y se marca si la respuesta es correcta o no (íconos ✅❌), y luego se debe pulsar el botón continuar (Fig. 7).

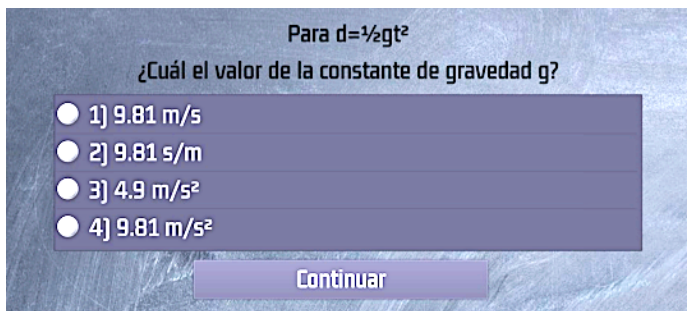

Figura 7. Pregunta temática aleatoria sobre caída libre y movimiento uniformemente acelerado.

De forma sintética, el algoritmo de juego (jugadas) es el siguiente: 1) se activa música de fondo e inicia tiempo de juego; 2) muestra personaje en posición inicial, se elege al azar un tiempo de caída ($t_{reto}$) y una posición del barril; 3) muestra reto y se coloca el barril; 4) usuario mueve peldaño con teclas cursor; 5) se espera evento de entrada de la tecla *enter* para iniciar *timer*, sonido y caída del objeto; 6) se espera evento de colisión objeto-peldaño para detener el *timer*, reproducir el sonido de choque y calcular el tiempo de caída; 7) se estima error para obtener la puntuación de la jugada; 8) se espera evento de colisión objeto-barril para activar sonido/efecto, indicar con una flecha la distancia correcta del peldaño para el tiempo-reto asignado y se actualiza la puntuación; 9) se espera evento de colisión objeto-piso para activar sonido/efecto, borrar personaje y regresar al paso 2; 10) se espera evento temporizado hasta agotar tiempo de juego para detener la jugada y cambiar al siguiente nivel.

### 3.3. Matematización y pruebas del fenómeno físico.

Uno de los procesos centrales para el desarrollo del juego fue, desde el punto de vista didáctico y formativo, la matemati–zación del fenómeno físico de caída libre de un objeto que parte del reposo y se desprecia la resistencia del aire. Para ello fue esencial establecer una simulación del fenómeno lo más eficiente y fidedigna posible para aproximar de forma correcta y precisa las magnitudes de las distancias y tiempos correspondientes al modelo matemático de la caída libre. Un factor clave fue corroborar si efectivamente el motor de física de Godot se ajustaba a dicho modelo. Durante las sesiones 7-10 (Tabla 1), se idearon los cálculos tiempo-distancia (matematizar), se realizaron las pruebas de medición de tiempos y los ajustes de calibración para empatar la escala precisa de la regleta de referencia (Fig. 6) para mapear correctamente las unidades de distancia basadas en píxeles de pantalla a las unidades métricas correspondientes, replicando las experimentaciones del plano inclinado de Galileo, se obtuvieron finalmente las mismas mediciones históricas, es decir, para t=1s la regla marcó d=1u, para t=2s se obtuvo d=4u y para t=3s fue d=9u (ver escala en Fig. 6). Como nota curiosa del motor de juego, se observó que cuando la computadora tenía muchas aplicaciones abiertas, para t=3s existía un pequeño desfase en la mayor distancia de hasta d=8.9u, por lo que se recomendó a los estudiantes que al obtener la constante de calibración cerrarán todas las aplicaciones posibles que estuvieran activas en sus computadoras personales. Gracias a la adecuada matematización del problema fue posible estimar correctamente en pixeles la distancia correspondiente para el tiempo-reto y estimar el error comparándola contra la distancia de impacto objeto-peldaño que seleccionó el jugador.

Matematizando formalmente el problema de caída libre de objetos, despreciando la resistencia del aire y partiendo del reposo, tenemos que la distancia recorrida viene dada por:

$$d = \tfrac{1}{2} g t^2$$

Dado que la intención del juego es descubrir dicha fórmula (aprendizaje inductivo), es posible reescribir dicha fórmula en dos partes, una parte constante y la otra en función del tiempo:

$$d = k \cdot f(t)$$

En el primer nivel del juego se pretende deducir solamente la relación de proporcionalidad, es decir $f(t)$, proporcionando un reto lúdico consistente en encontrar la distancia recorrida para un tiempo-reto aleatorio asignado ($t_{reto}$). El jugador debe estimar, en fracciones de segundo, cómo deberá desplazar el peldaño mediante el cursor hasta la distancia donde deberá impactar objeto-peldaño en $t_{reto}$. A la vez, debe girar el peldaño para lograr que el objeto rebote y caiga dentro del barril ($B_{hit}$), logrando así una mayor puntuación. Al finalizar la jugada aparecerá una flecha en la regleta indicando la distacia correcta. De esta forma es posible partir de cero conocimientos y mediante esta mecánica de juego, ir reforzando el aprendizaje hasta lograr aproximar un modelo mental más próximo al modelo físico correcto. El tiempo de impacto objeto-peldaño ($t_{hit}$) y el tiempo del reto ($t_{reto}$) son utilizados para estimar el error y con ello calcular la puntuación de cada jugada (máx. 100 pts), según:

$$score = 50 \cdot \left(1 - \frac{|t_{reto}-t_{hit}|}{t_{max}}\right) + 50 \cdot B_{hit}$$

Con esta dinámica, luego de varias jugadas se espera que el jugador deduzca $f(t)$, por ejemplo para $t_{reto}$=[1,2,3], un jugador que sólo observe y recuerde d=[2,4], deducirá de forma incorrecta $f(t) = 2t$. Sin embargo, si el jugador observó o logró los resultados d=[2,4,9], podra acertar a deducir que dicha parte de la fórmula corresponde a $f(t) = t^2$.

De acuerdo con el algoritmo de juego presentado arriba, al agotarse el tiempo del primer nivel, aparecerá una pregunta al azar para corroborar el conocimiento adquirido (Fig. 7). En caso de que el jugador falle, lo cual indica que aún no puede inferir correctamente la relación de proporcionalidad tiempo-distancia, regresará al primer nivel para reforzar una experiencia más exitosa y lograr comprender dicha propiedad del fenómeno. Si contesta correctamente, avanza al siguiente nivel donde debe deducir la magnitud de la constante $k$, y por consecuencia, la constante de la aceleración gravitacional $g$.

De forma análoga al nivel precedente, se genera una distancia aleatoria ($d_{reto}$) y el jugador debe mover el peldaño para aproximarlo a la distancia dada en unidades métricas (Fig. 6). Al agotarse el tiempo, una pregunta al azar verifica el conocimiento logrado (Fig. 7). Por ejemplo, para $d_{reto}$=4.9m, deberá ubicar la repisa en $d_u$=1u (regleta), correspondiendo a $t_{hit}$=1s, es decir:

$$d = \frac{g}{2}t^2 = k \cdot f(1) = k \cdot 1s^2 = 4.9m,$$

$$\text{o sea } k = \frac{d}{t^2} = \frac{4.9m}{1s^2} = 4.9m/s^2,$$

$$\text{y por lo tanto, } g = 9.8m/s^2.$$

## 4. RESULTADOS

Gracias a la selección de un entorno y motor de videojuegos fue posible lograr como resultado que los estudiantes, sin contar experiencia previa de programación lograran ser capacitados en 9 semanas (sesiones 3-11, Tabla 1) para conocer todos los elementos básicos necesarios para crear el videojuego. Desde luego, la confluencia de los diferentes métodos permitió derivar una metodología tanto didáctica como ágil para el desarrollo del software del videojuego (Fig. 3). De esta forma se logró que todos los 10 estudiantes del curso de Física, sin excepción, lograran terminar su videojuego por completo y sin errores, que en promedio, el juego terminado requirió un total de 7 scripts, donde, 3 scripts ocuparon menos de 10 líneas de código y los 4 scripts restantes no rebasaron las 100 líneas de código, incluyendo comentarios y espacios dentro de los programas, lo que arrojó un total de 250 líneas de código (sin renglones en blanco ni comentarios),

## 5. TRABAJO A FUTURO

Para futuras versiones dado que pero cabe destacar que dado que la lógica de juego de los cuatro niveles de juego es muy similar, dos scripts muy parecidas para la simulación de los niveles 1 y 3, y la misma lógica de ambos cuestionarios de los niveles 2 y 4 (solo cambia el banco de preguntas), es muy viable reducir hasta un 20% el código requerido para obtener la misma funcionalidad. Sin embargo, esto requiriría un nivel más avanzado de programación, por lo que tal vez no sea recomendable complicar demasiado el programa para hacerlo más accesible a estudiantes de primeros semestres de ingeniería que aún no han usado un lenguaje de programación. También es posible incorporar, siguiensdo la misma metodología mas temas revisados en el curso de Física, como la dinámica de fuerzas manejando vectores, la inercia, fricción, movimiento parabólico y circular, etc. Algo que además soporta el entorno de Godot y resultaría ser muy práctico y atractivo sería migrar el juego a una plataforma móvil para que una vez concluido se pudiera descargar y correr directamente en un teléfono inteligente.

## 6. CONCLUSIONES

A diferencia de usar un juego o paquete educativo para Física (aprendizaje lúdico pasivo), se decidió explorar la alternativa mas arriesgada de crear un juego serio, para que el estudiante demuestre su creatividad y capacidad de aplicar los conceptos teóricos vistos en clase. Esto se logró gracias a la combinación de diversos métodos y estrategias didácticas (aprendizaje activo basado en proyectos, mentorías, retos desvanecidos, teoría cognitiva, taxonomía de Bloom, método de Halsted y otros descritos en (Fig. 3), en conjunto con una metodología y herramienta de desarrollo apropiada (programación en vivo, pensamiento computacional, protototipos, desarrollo ágil en espiral). Sin embargo, es importante, en lo que respecta al presente artículo, el gran impacto e importancia del proceso de matematización de la Física, ya que el problema mas crítico fue representar e implementar un modelo matemático fiel y preciso del fenómeno físico hasta un nivel *pixel perfect* de alta eficiencia y precisión. Por lo tanto, la sinergía lograda entre el pensamiento matemático y el pensamiento computacional permitió coadyuvar a que el estudiante pasará del ámbito teórico de la Física al campo aplicado de la Física de una forma activa y experiencial que permitió consolidar sus conocimientos, manifestando su creatividad y desarrollando otras habilidades digitales de gran auge en la actualidad (edición de imágenes, audio, video y la programación computacional).

Si bien el método de aprendizaje activo supervisado con retos se probó experimentalmente y se reportó en [16], es importante mencionar que se efecturon tanto encuesta de inicio como al final del curso, donde se detectó que la mayor parte de los estudiantes consideró que mejoró su actitud, conocimientos y experiencias tanto a nivel teórico como de aplicación de la Física, en esta parte del trabajo enfocada al desarrollo del software del juego serio,

Es importante destacar entre los hallazgos del experimento didáctico, reflexionando en torno a los factores exitosos, esta desde luego el factor motivacional y lúdico positivo propio de un videojuego, pero desde el punto de vista del enfoque metódico y didáctico, resulto más asequible y apropiado un enfoque ingenieril y aplicativo de los temas teóricos de la Física.

**Agradecimientos.**

**Referencias.**